\documentclass[conference]{IEEEtran}
\IEEEoverridecommandlockouts
\usepackage{cite}
\usepackage{amsmath,amssymb,amsfonts}
\usepackage{algorithmic}
\usepackage{graphicx}
\usepackage{textcomp}
\usepackage{xcolor}
\def\BibTeX{{\rm B\kern-.05em{\sc i\kern-.025em b}\kern-.08em
    T\kern-.1667em\lower.7ex\hbox{E}\kern-.125emX}}
\begin{document}

\title{PAS-QFL: Personalized Ansatz Selection for Quantum Federated Learning under Client Data Heterogeneity
}

\author{\IEEEauthorblockN{Jindi Wu}
\IEEEauthorblockA{\textit{School of computing} \\
\textit{DePaul University}\\
Chicago, USA \\
jwu115@depaul.edu}
\and
\IEEEauthorblockN{Qun Li}
\IEEEauthorblockA{\textit{Computer science department} \\
\textit{William \& Mary}\\
Williamsburg, USA \\
liqun@cs.wm.edu}
}

\maketitle

\begin{abstract}
Quantum federated learning (QFL) lets multiple quantum clients collaboratively train quantum neural networks (QNNs) without sharing private local data. However, existing QFL methods commonly assume that all clients use the same ansatz, overlooking how heterogeneous client data affects ansatz suitability. Under class-imbalanced non-IID data, different clients may favor different ansatz structures, so a fixed ansatz can lead to unstable and unfair performance across clients. In this paper, we propose PAS-QFL, a Personalized Ansatz Selection framework for QFL under client data heterogeneity. Rather than treating the ansatz as a monolithic structure, PAS-QFL decomposes each client QNN into a globally shared ansatz and a client-specific private ansatz, and personalizes the structure of the private ansatz rather than only its parameters. The shared ansatz is placed first and selected by a stability-aware cross-client criterion so that its parameters can be reliably aggregated, while the private ansatz serves as a personalized decision head, selected per client by local Macro-F1 to adapt the shared representation to its local data. During training, each client updates both its shared and private parameters locally but uploads only the shared parameters, so federated aggregation stays well-defined while each client keeps its own private structure. PAS-QFL uses Macro-F1 as the primary selection metric to avoid misleading accuracy under class imbalance. Experiments on heterogeneous QFL tasks show that PAS-QFL improves average Macro-F1 over the existing fixed-ansatz QFL baselines, demonstrating the value of personalizing the ansatz structure for practical QFL.
\end{abstract}


\section{Introduction} \label{sec:intro}

Quantum-networked systems are emerging as an important infrastructure for connecting distributed quantum devices~\cite{azuma2023quantum}. In near-term deployments, these quantum clients are expected to access local noisy intermediate-scale quantum (NISQ) processors while coordinating primarily through classical communication links. Because their local data is privacy-sensitive and cannot be freely shared across the network, this setting naturally motivates quantum federated learning (QFL) \cite{xia2021quantumfed}, where distributed quantum clients collaboratively train quantum neural networks (QNNs) by exchanging classical model parameters rather than raw local data~\cite{ren2025toward}. QNNs are typically realized as variational quantum circuits (VQCs), whose ansatz determines the circuit structure, entanglement pattern, and semantics of the trainable parameters~\cite{farhi2018classification}. In QFL, the ansatz plays two coupled roles: it shapes the quantum representation learned from local data, and it defines the parameterized decision structure whose parameters are aggregated across clients. Therefore, ansatz design is not only a local modeling choice, but also a federated collaboration choice.

\begin{figure}[t]
\centering
\includegraphics[width=\linewidth]{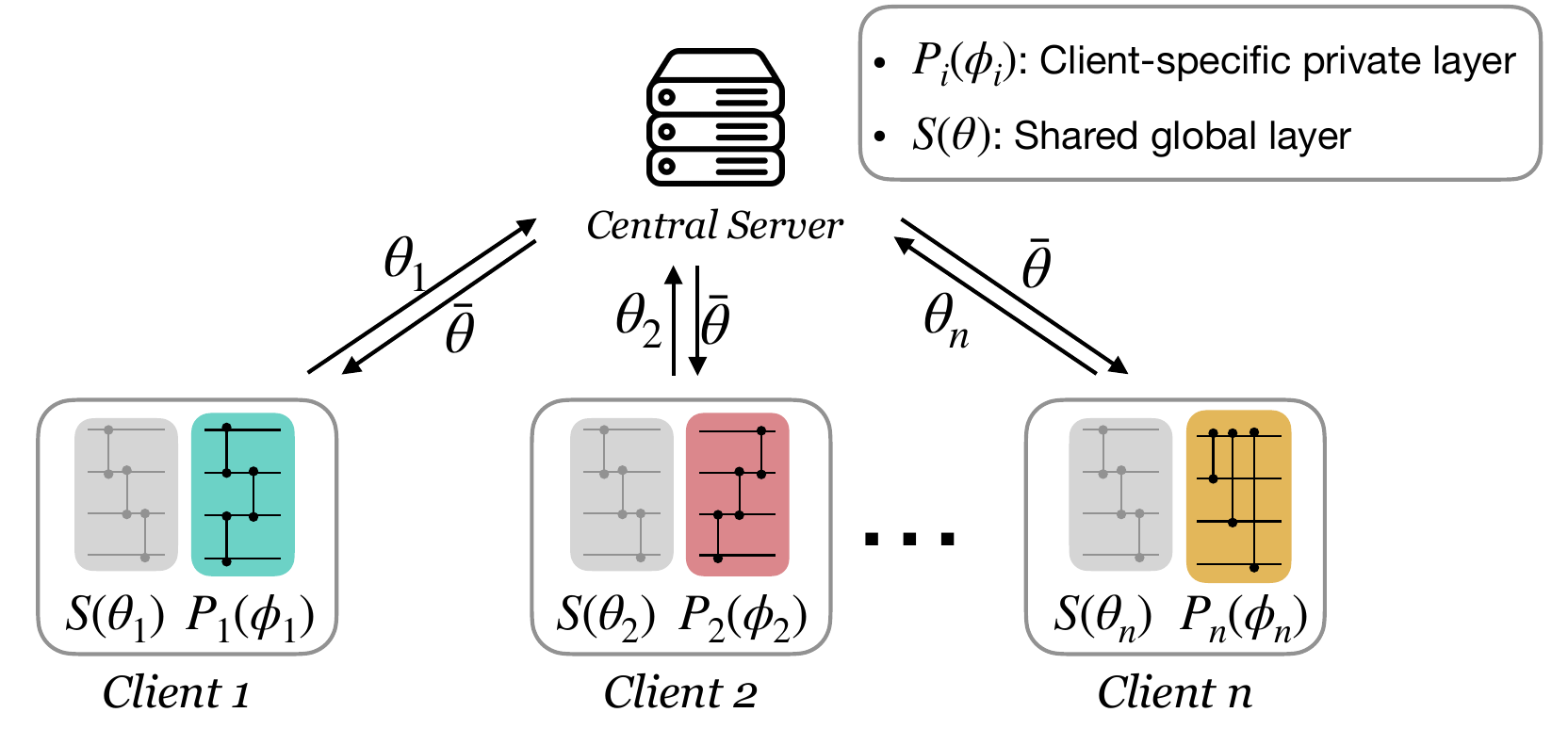}
\caption{\textbf{Overview of PAS-QFL.}  Each client QNN is decomposed into a globally shared ansatz $S(\theta_i)$, and a client-specific private ansatz $P_i(\phi_i)$ that follows it as a personalized decision head. The shared ansatz is selected for stable cross-client performance, while each private ansatz is selected per client by local profiling. During federated training, every client updates both its private parameters $\phi_i$ and its shared parameters $\theta_i$ on local data, but uploads only $\theta_i$ to the central server, which aggregates them into $\bar{\theta}$ and broadcasts $\bar{\theta}$ back for the next round. The private ansatz structure and its
parameters remain local and are never shared.}
\label{fig:overview}
\end{figure}

A central challenge in QFL is client data heterogeneity. In networked deployments, clients often collect data from different heterogeneous environments, resulting in non-independent and identically distributed (non-IID) local datasets~\cite{zhao2018federated}. However, most existing QFL methods adopt a homogeneous design, where all clients use the same ansatz and aggregate the same set of parameters through a central server~\cite{mcmahan2017communication, chen2021federated, wu2024distributed}. This design simplifies aggregation, but it also forces heterogeneous clients to share the same quantum representation and decision structure. Under class-imbalanced non-IID data, such a fixed ansatz may favor some clients while limiting others, leading to unstable and unfair client-level performance.

We confirm this issue through a motivating profiling study in Section~\ref{sec:motivation}, where each client locally evaluates a pool of ansatzes with different entanglement structures. The resulting Macro-F1 profiles differ markedly across clients, and no single ansatz consistently dominates all clients. This indicates that ansatz suitability is client-dependent under class-imbalanced non-IID data. A direct solution would be to let each client adopt its own best full ansatz. However, corresponding parameters may have different semantic meanings across clients, rendering direct parameter averaging inappropriate.
Thus, the key challenge is to support client-specific architectural adaptation while preserving a common component whose parameters remain semantically consistent across clients.

Personalized QFL provides a natural starting point by allowing each client to maintain a private layer in addition to a shared federated component. Nevertheless, most personalized designs still keep the private layer architecture fixed across clients and personalize only its parameters. This limits the adaptation capacity of the private component. For VQCs, the entanglement topology and gate layout define circuit-level inductive biases that may not be fully compensated by parameter tuning within a single fixed ansatz. Our profiling study suggests that clients with different class ratios can prefer different ansatz structures, motivating personalization not only at the parameter level but also at the ansatz level. More fundamentally, the shared and private components serve distinct roles: the shared component should support stable cross-client aggregation, while the private component should adapt the shared model to each client's local decision behavior. These observations motivate selecting the shared and private ansatzes separately according to their different roles.

We propose \textbf{PAS-QFL}, a \textbf{P}ersonalized \textbf{A}nsatz \textbf{S}election framework for \textbf{QFL} under client data heterogeneity. PAS-QFL decomposes each client QNN into a globally shared ansatz and a client-specific private ansatz, as shown in Fig.~\ref{fig:overview}. The shared ansatz is placed before the private component and is selected to provide stable performance across clients, so that its parameters can be meaningfully aggregated through federated training. The private ansatz is placed after the shared component and serves as a personalized quantum decision head, adapting the shared representation to each client's imbalanced local data. During training, each client updates both the shared parameters and its private parameters, but only the shared parameters are uploaded and aggregated. By personalizing the structure of the private layer rather than only its parameters, PAS-QFL enables stronger client-level adaptation while preserving well-defined federated aggregation. 
This paper makes the following main contributions.
\begin{itemize}
\item We identify client-dependent ansatz suitability as a key source of performance instability in QFL under non-IID data, showing that no single ansatz consistently performs best across heterogeneous clients.
\item We propose PAS-QFL, a personalized ansatz selection framework that decomposes each client QNN into a globally shared ansatz and a client-specific private ansatz. 
\item We evaluate PAS-QFL against homogeneous and fixed-private QFL baselines under non-IID data, demonstrating its consistent performance advantages.
\end{itemize}

\section{Related Work}

Several recent works address client heterogeneity in QFL at the aggregation or parameter level. Weighted aggregation methods assign client-specific aggregation weights according to the distribution, reliability, or quality of local data~\cite{gurung2025performance}, while personalized approaches introduce regularized local updates so that each client can adapt its parameters to its own data and training behavior while remaining consistent with the global model~\cite{rahman2025toward}. These methods improve aggregation robustness and client-level parameter adaptation, but they typically retain a fixed ansatz architecture. As a result, they do not directly address how ansatz-induced decision behavior affects different clients under heterogeneous local data.

Other works consider structural heterogeneity through shared-local QFL designs. In these methods, a shared global component participates in federated aggregation, while a personalized local component remains private to each client~\cite{shi2024personalized}. Quorus further studies heterogeneous quantum clients with different hardware capabilities and supports varying circuit depths through layerwise losses~\cite{han2025layerwise}. These studies demonstrate the value of personalization and model heterogeneity in QFL. However, the private component is usually predefined, or the structural variation is driven by hardware constraints rather than by each client’s local data behavior. In contrast, PAS-QFL treats the private-layer ansatz structure itself as a personalization target. It selects a client-specific private ansatz as a quantum decision head for local adaptation, while separately selecting a shared ansatz that remains structurally consistent for stable federated aggregation. Although other factors, such as quantum noise \cite{wu2026fine}, may also affect quantum ansatz selection, we focus specifically on local data heterogeneity in the QFL setting.

\section{Motivation and Problem Statement} \label{sec:motivation}

\begin{figure}[t]
\centering
\includegraphics[width=\linewidth]{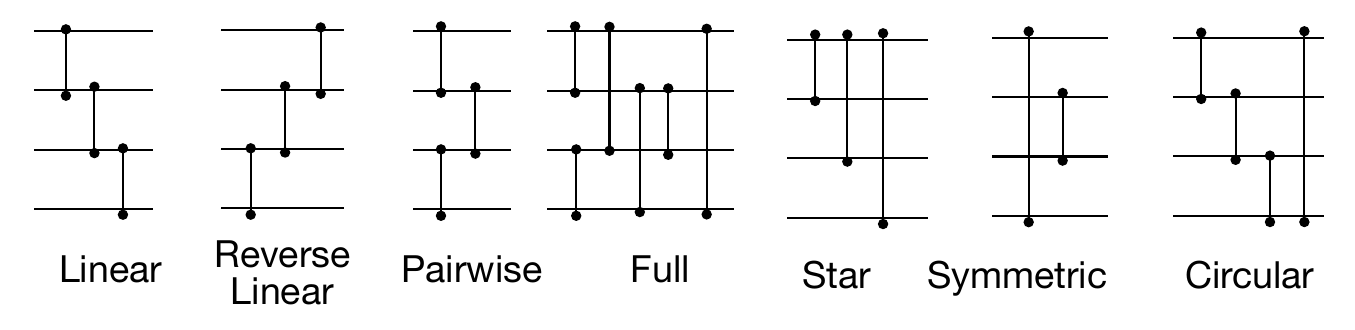}
\caption{Candidate ansatz pool. PAS-QFL considers multiple ansatz structures with different entanglement patterns.}
\label{fig:pool}
\end{figure}

To understand how ansatz structure affects heterogeneous QFL clients, we conduct a local ansatz profiling study. We consider a pool of commonly used ansatzes (Fig.~\ref{fig:pool}), which use the same number of qubits but differ in their entanglement structures, including linear, reverse-linear, pairwise, full, star, symmetric, and circular patterns. We simulate $N=9$ clients, each building a 5-qubit QNN and receiving 400 local training samples from a binary classification task derived from Fashion-MNIST. All clients have the same local training size but different class ratios, varying gradually from 1:9 to 9:1. Thus, each client observes an imbalanced local distribution, while the pooled data across all clients remain class-balanced. On each client, every candidate ansatz is trained as a standalone local QNN and evaluated using Macro-F1 on a balanced validation set. Since the private component in PAS-QFL serves as a client-specific quantum decision head, this local profiling provides a natural proxy for how suitable each ansatz is for local decision adaptation.

\begin{figure}[t]
\centering
\includegraphics[width=\linewidth]{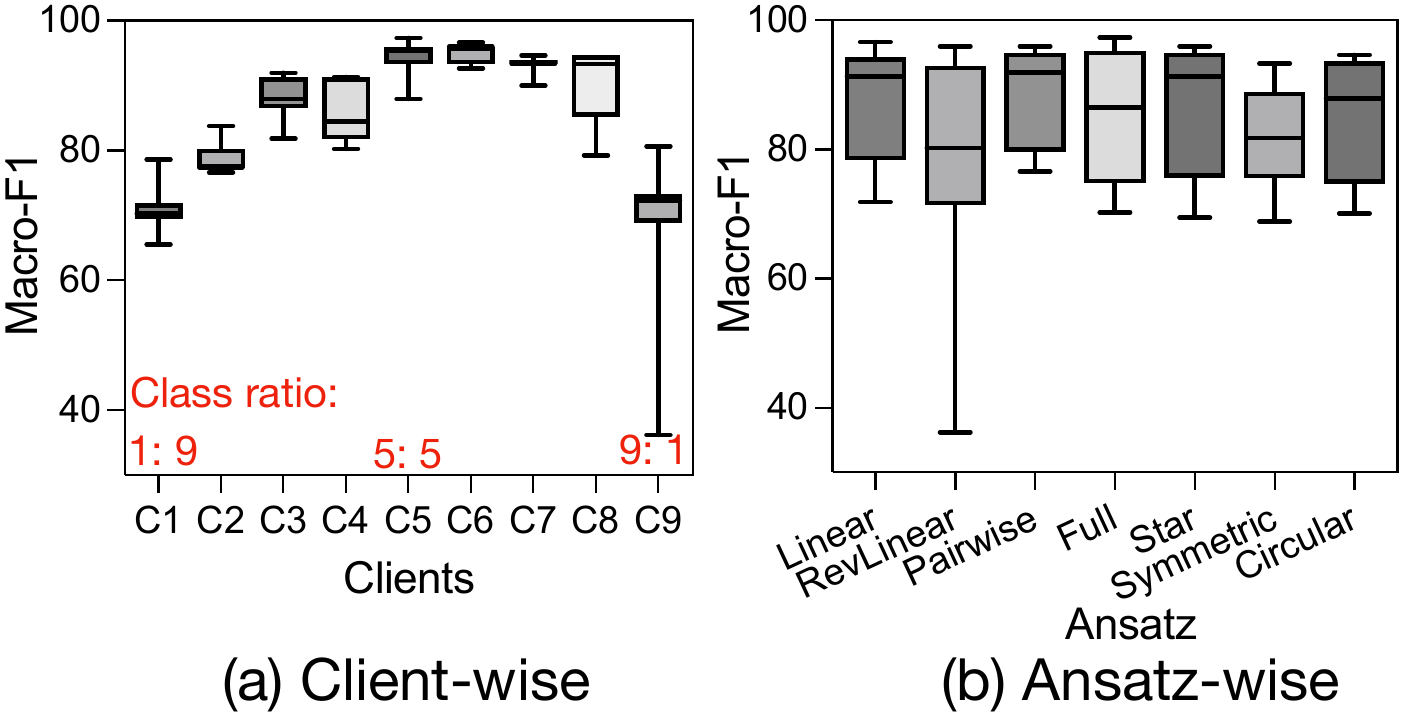}
\caption{Local ansatz profiling under client data heterogeneity. The client-wise view summarizes the validation Macro-F1 distribution of different ansatzes on each client, showing client-dependent sensitivity to ansatz choice. The ansatz-wise view summarizes the validation Macro-F1 distribution of each ansatz across clients, showing different levels of their cross-client stability.}
\label{fig:single_local}
\end{figure}

Fig.~\ref{fig:single_local} summarizes the profiling results from two complementary views. The client-wise view in Fig.~\ref{fig:single_local}(a) reports, for each client, the distribution of validation Macro-F1 scores across candidate ansatzes. The spread varies substantially across clients: some clients are relatively insensitive to ansatz choice, while others show a large gap between their best and worst ansatzes. This indicates that ansatz suitability is client-dependent under class-imbalanced non-IID data. Therefore, using the same private-layer ansatz structure for all clients may fail to capture client-specific decision behavior.

The ansatz-wise view in Fig.~\ref{fig:single_local}(b) reports, for each ansatz, the distribution of validation Macro-F1 scores across clients. Different ansatzes exhibit different levels of cross-client stability. Some ansatzes perform consistently across clients, while others achieve strong performance on certain clients but degrade on others. This suggests that the shared ansatz should not be selected only by average performance, but should also account for stability across heterogeneous clients, since its parameters are the ones aggregated during federated training.

These observations motivate a role-specific ansatz selection problem. Let $\mathcal{A}=\{a_1,\dots,a_M\}$ be the candidate ansatz pool, and consider $N$ clients with heterogeneous local datasets $\{D_1,\dots,D_N\}$. PAS-QFL aims to select a common shared ansatz $S\in\mathcal{A}$ and a client-specific private ansatz $P_i\in\mathcal{A}$ for each client $i$. 
The goal is to achieve high average Macro-F1 across heterogeneous clients, while keeping the shared ansatz structurally identical so that aggregation remains well-defined.
Exhaustively training all possible shared-private configurations would require evaluating a large number of combinations, which is infeasible even for small ansatz pools. The central challenge is therefore to perform ansatz selection efficiently while accounting for both client-specific adaptation and cross-client stability.

\section{PAS-QFL Design}

\textbf{Overview.}
PAS-QFL personalizes QFL at the ansatz-architecture level while preserving a shared component for federated aggregation. Instead of forcing all clients to use the same QNN ansatz, PAS-QFL decomposes each client model into a globally shared ansatz and a client-specific private ansatz. For client $i$, the QNN circuit is defined as
\begin{equation} \label{eq:qnn}
    P_i(\phi_i)\circ S(\theta_i)\circ U_{\mathrm{enc}}(x),
\end{equation}
where $U_{\mathrm{enc}}(x)$ encodes the input data $x$, $S(\theta_i)$ is client $i$'s local copy of the shared ansatz, and $P_i(\phi_i)$ is the client-specific private ansatz. The shared ansatz is placed before the private component and provides a common federated block whose parameters can be aggregated across clients. The private ansatz is placed after the shared component and serves as a personalized quantum decision head for local adaptation.

PAS-QFL proceeds in four stages. First, each client performs local ansatz profiling: for every candidate ansatz in the pool, client $i$ trains a standalone local QNN and records its validation Macro-F1. This produces a client-ansatz profiling matrix that captures how each ansatz behaves on each heterogeneous client. Second, PAS-QFL selects the personalized private ansatz for each client from the client-wise profiling results. The selected $P_i(\phi_i)$ serves as a client-specific quantum decision head and remains local to client $i$. Third, PAS-QFL selects the shared ansatz from the ansatz-wise profiling results using a stability-aware criterion that considers both average Macro-F1 and cross-client variance. The selected $S(\theta_i)$ is structurally identical across clients so that its parameters can be aggregated. Finally, PAS-QFL performs federated training with the selected architecture $P_i(\phi_i)\circ S(\theta_i)\circ U_{\mathrm{enc}}(x)$. During each communication round, client $i$ updates both $\phi_i$ and $\theta_i$ on local data, but uploads only $\theta_i$ to the server. The server aggregates the uploaded shared parameters into $\bar{\theta}$ and broadcasts them back to clients, while the private ansatz structures and parameters remain local.

\subsection{Client-Specific Private Ansatz Selection}

PAS-QFL first selects a personalized private ansatz for each client from the local profiling results. Let $F_i(a)$ denote the validation Macro-F1 obtained when client $i$ trains candidate ansatz $a\in\mathcal{A}$ as a standalone local QNN under the profiling budget. Since the private component in PAS-QFL serves as a client-specific quantum decision head, a high $F_i(a)$ indicates that the circuit structure and entanglement pattern of $a$ are well matched to client $i$'s local decision behavior. PAS-QFL therefore selects the private ansatz of client $i$ as
\begin{equation} \label{eq:local}
P_i = \arg\max_{a\in\mathcal{A}} F_i(a).
\end{equation}
Because clients have heterogeneous and class-imbalanced local data, the selected private ansatz $P_i$ may differ across clients. This allows PAS-QFL to personalize not only the private parameters but also the private-layer architecture itself. The selected $P_i(\phi_i)$ is placed after the shared ansatz and remains local to client $i$, as shown in Eq.~\ref{eq:qnn}.
During federated training, the private parameters $\phi_i$ are updated locally together with the client copy of the shared parameters, but neither the private structure nor the private parameters are uploaded or aggregated. This preserves client-specific architectural adaptation while keeping aggregation defined only over the common shared ansatz.

\subsection{Stability-Aware Shared Ansatz Selection}

PAS-QFL next selects a single shared ansatz used by all clients. Unlike the private ansatz, whose role is to adapt to each client's local decision behavior, the shared ansatz must support stable cross-client collaboration because its parameters are aggregated during federated training. Therefore, the shared ansatz should not be selected only by its performance on a few clients, but by its overall performance and stability across heterogeneous clients.

Using the same profiling matrix, PAS-QFL scores each candidate ansatz $a$ by its cross-client mean Macro-F1 penalized by its cross-client spread:
\begin{equation} \label{eq:score}
\mathrm{Score}(a) = \mu(a) - \lambda \sigma(a),
\end{equation}
where $\mu(a)=\frac{1}{N}\sum_{i=1}^{N}F_i(a)$
and $\sigma(a)$ is the standard deviation of $\{F_i(a)\}_{i=1}^{N}$ across clients. PAS-QFL selects the shared ansatz as
\begin{equation}
S = \arg\max_{a\in\mathcal{A}}\mathrm{Score}(a)
\end{equation}
The coefficient $\lambda\ge 0$ controls the trade-off between average performance and cross-client stability. When $\lambda=0$, PAS-QFL selects the ansatz with the highest mean Macro-F1; larger $\lambda$ penalizes ansatzes whose performance varies substantially across clients. 

\subsection{Federated Training}

After the shared ansatz $S$ and the client-specific private ansatzes $\{P_i\}_{i=1}^{N}$ are selected, PAS-QFL performs federated training over the shared component while keeping each private component local. The selected ansatz structures remain fixed during training, but their parameters are optimized. For client $i$, the model is shown as Eq.~\ref{eq:qnn}.


At the beginning of communication round $t$, the server broadcasts the current global shared parameters $\bar{\theta}^{t}$ to all clients. Each client initializes its local shared parameters as $\theta_i^{t}=\bar{\theta}^{t}$ and trains the full model on its local data, updating both the shared parameters and its private-head parameters:
\begin{equation}
(\theta_i^{t+1}, \phi_i^{t+1})
\leftarrow
\mathrm{LocalTrain}(\theta_i^t,\phi_i^t;D_i).
\end{equation}
After local training, client $i$ uploads only the updated shared parameters $\theta_i^{t+1}$ to the server, while the private ansatz $P_i$ and its parameters $\phi_i^{t+1}$ remain local and are never transmitted. The server aggregates the uploaded shared parameters using FedAvg:
\begin{equation}
\bar{\theta}^{t+1}
=
\sum_{i=1}^{N}
\frac{n_i}{\sum_{j=1}^{N} n_j}
\theta_i^{t+1},
\end{equation}
where $n_i$ is the number of local training samples on client $i$. When all clients have the same local training size, this reduces to equal-weight averaging.

Because every uploaded parameter belongs to the same shared ansatz $S$, federated aggregation remains well-defined. At the same time, clients can use different private ansatz structures as personalized quantum decision heads, since these private structures and parameters are kept local and are never aggregated.

\section{Evaluation}

We evaluate PAS-QFL using Qiskit-based quantum circuit simulations under both noiseless and noisy settings. The noiseless simulation isolates the impact of ansatz selection under client data heterogeneity, while the noisy simulation examines whether the design remains effective under NISQ noise. We consider QFL systems with 5 and 9 clients, where the local class ratios vary from $1{:}9$ to
$9{:}1$ to create class-imbalanced non-IID data. Each client builds a $5$-qubit
QNN and holds $125$ training samples per class, with validation and test sets
each sized at $20\%$ of the training set. For all experiments, we use angle
encoding and the candidate ansatz pool in Fig.~\ref{fig:pool}, and all qubits
are measured to produce the predicted class. Parameters are optimized with
simultaneous perturbation stochastic approximation (SPSA); each client performs
$20$ local epochs per communication round, and training runs for $30$
communication rounds. For shared ansatz selection, we evaluate
$\lambda \in \{0.1, 0.3, 0.5\}$ and set $\lambda = 0.1$ in Eq.~\eqref{eq:score}.
We evaluate PAS-QFL on the datasets Fashion-MNIST \cite{xiao2017fashion} and MNIST \cite{deng2012mnist}, considering both binary and four-class classification tasks:
\begin{itemize}
\item \textbf{FM-2}: binary classification on Fashion-MNIST.
\item \textbf{FM-4}: four-class classification on Fashion-MNIST.
\item \textbf{M-2}: binary classification on MNIST.
\item \textbf{M-4}: four-class classification on MNIST.
\end{itemize}
We compare four training settings:
\begin{itemize}
\item \textbf{Local}: Each client trains its best QNN independently on its local data without federated aggregation. This setting evaluates the performance of purely local learning.
\item \textbf{Fix-all}: All clients use the same fixed ansatz structure, and all trainable parameters are shared and aggregated through federated learning. This represents conventional homogeneous QFL.
\item \textbf{Fix-pri}: Each client uses a shared-private QNN structure, where the private layer remains local but its ansatz structure is fixed and identical across clients. This setting represents parameter-level personalization with a fixed private-layer ansatz.
\item \textbf{PAS-QFL}: Each client uses the shared ansatz selected by the stability-aware criterion according to Eq.~\ref{eq:score} and a client-specific private ansatz selected according to Eq.~\ref{eq:local}. Only the shared parameters are uploaded and aggregated, while the personalized private ansatz and its parameters remain local.
\end{itemize}

\subsection{Overall Performance}

\begin{figure}[t]
\centering
\includegraphics[width=\linewidth]{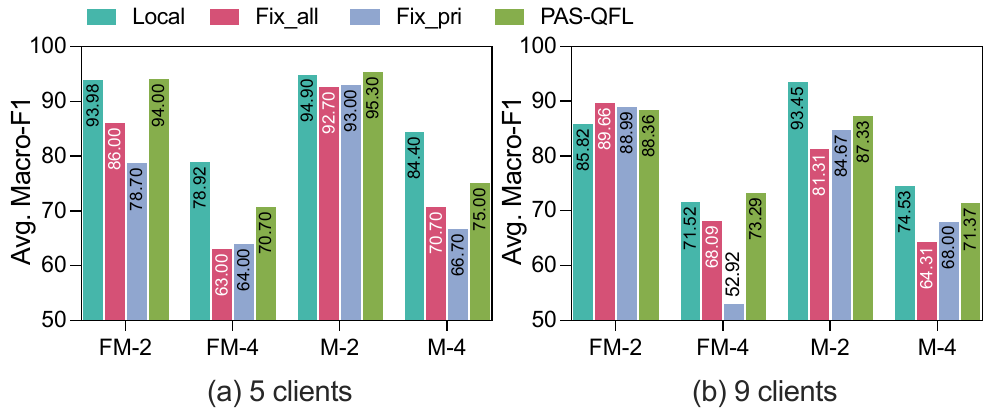}
\caption{Overall performance under noiseless simulation.}
\label{fig:5client_qfl}
\end{figure}

Fig.~\ref{fig:5client_qfl} reports the overall average Macro-F1 under noiseless simulation. PAS-QFL achieves strong and competitive performance across different datasets, task complexities, and client scales. In the 5-client setting, PAS-QFL outperforms both federated fixed-structure baselines, \textsc{Fix-all} and \textsc{Fix-pri}, on all four tasks. For example, on FM-2, PAS-QFL improves the average Macro-F1 from 86.00 and 78.70 to 94.00, and on M-4, it improves Macro-F1 from 70.70 and 66.70 to 75.00. These results show that using a client-specific private ansatz structure provides additional adaptation benefits beyond homogeneous QFL and fixed-private personalized QFL.

In the 9-client setting, PAS-QFL remains competitive on FM-2 and achieves stronger performance on the more challenging FM-4, M-2, and M-4 tasks compared with the fixed-structure QFL baselines. The improvement is especially clear on FM-4, where PAS-QFL increases the average Macro-F1 from 68.09 for \textsc{Fix-all} and 52.92 for \textsc{Fix-pri} to 73.29. This suggests that as client heterogeneity and task complexity increase, fixing the private-layer structure can become insufficient, while selecting client-specific private ansatzes helps improve local adaptation.

\begin{figure}[t]
\centering
\includegraphics[width=\linewidth]{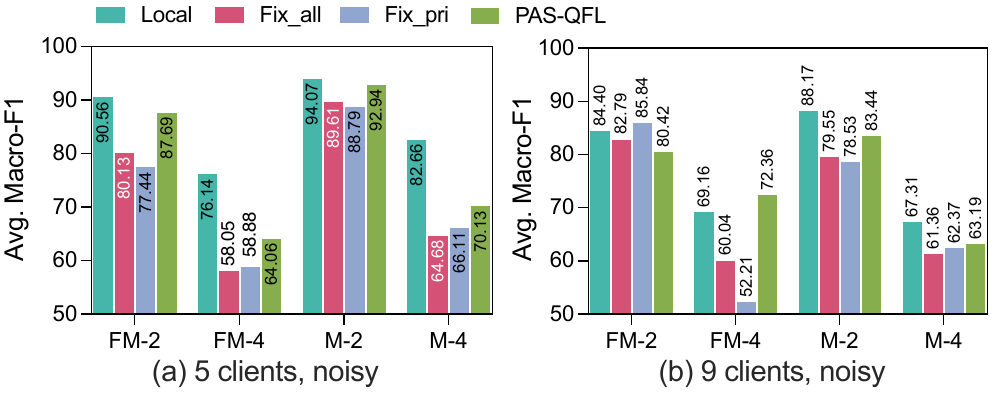}
\caption{Overall performance under noisy simulation.}
\label{fig:5client_qfl_noisy}
\end{figure}

Fig.~\ref{fig:5client_qfl_noisy} shows the corresponding results under noisy simulation. As expected, circuit noise generally reduces the average Macro-F1 of all methods. Nevertheless, PAS-QFL continues to outperform the two fixed-structure federated baselines in most settings. In the 5-client noisy setting, PAS-QFL improves over both \textsc{Fix-all} and \textsc{Fix-pri} on all four tasks. For example, on FM-4, PAS-QFL achieves 64.06 average Macro-F1, compared with 58.05 for \textsc{Fix-all} and 58.88 for \textsc{Fix-pri}; on M-4, PAS-QFL improves the score to 70.13, compared with 64.68 and 66.11. In the 9-client noisy setting, PAS-QFL is lower than the fixed-structure baselines on FM-2, but it outperforms them on FM-4, M-2, and M-4. The largest gain again appears on FM-4, where PAS-QFL achieves 72.36 average Macro-F1, substantially higher than 60.04 for \textsc{Fix-all} and 52.21 for \textsc{Fix-pri}.

On the simpler FM-2 task with 9 noisy clients, the additional private-head flexibility does not translate into higher average Macro-F1, possibly because binary classification is already well captured by a stable fixed structure and noisy local updates introduce additional variance. However, on more complex multi-class tasks, PAS-QFL consistently provides larger gains.

Overall, these results support the main design motivation of PAS-QFL. Homogeneous QFL and fixed-private personalized QFL can be effective in some simple settings, but their fixed ansatz structures limit their ability to adapt to heterogeneous clients. PAS-QFL improves this by selecting a stable shared ansatz for aggregation and client-specific private ansatzes for local decision adaptation. The benefit is most evident on more challenging multi-class tasks and remains visible under noisy simulation, indicating that ansatz-level personalization is useful not only in ideal circuit execution but also under NISQ noise.

\subsection{Effect of Personalized Private Ansatz Selection}

\begin{figure}[t]
\centering
\includegraphics[width=0.8\linewidth]{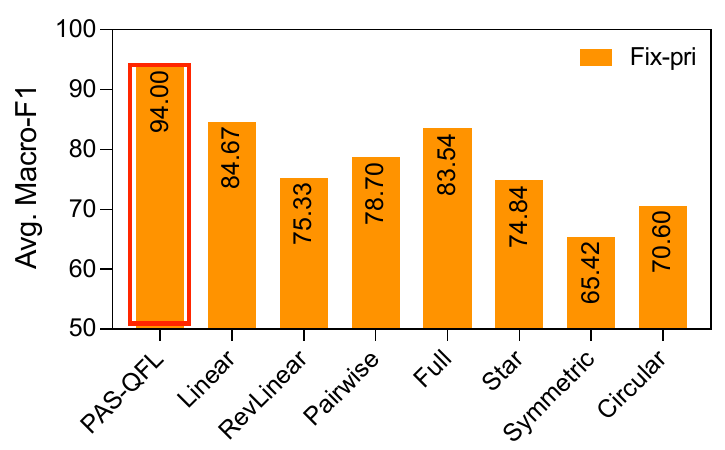}
\caption{Effect of personalized private ansatz selection on the FM-2 task with 5 clients. Fixed-private baselines use the same private ansatz structure for all clients, while PAS-QFL selects client-specific private ansatzes through local profiling.}
\label{fig:9c_5q_private_selectin}
\end{figure}

We further evaluate whether the improvement of PAS-QFL comes from personalizing the private ansatz structure. We focus on the FM-2 task with 5 clients and compare PAS-QFL with a set of fixed-private baselines. Pairwise ansatz is selected as the shared layer. In each Fix-pri baseline, all clients use the same private ansatz structure, while the private parameters remain locally trainable. Therefore, these baselines represent parameter-level personalization with a fixed private-layer architecture.

Fig.~\ref{fig:9c_5q_private_selectin} shows that PAS-QFL achieves the highest average Macro-F1 of 94.00, outperforming all fixed-private variants. The best fixed-private baseline uses the linear ansatz and reaches 84.67, while other fixed private structures obtain lower performance, ranging from 65.42 to 83.54. This large performance gap among fixed-private baselines indicates that the private ansatz structure has a substantial impact on client-level adaptation. More importantly, no single fixed private ansatz matches PAS-QFL, which selects the private ansatz separately for each client according to local profiling.

These results demonstrate that the benefit of PAS-QFL does not merely come from introducing a private layer. Instead, the structure of the private layer matters. By allowing different clients to use different private ansatz structures as personalized quantum decision heads, PAS-QFL provides stronger adaptation to heterogeneous local data than fixed-private personalized QFL.

\subsection{Effect of Stability-Aware Shared Ansatz Selection}

\begin{figure}[t]
\centering
\includegraphics[width=0.78\linewidth]{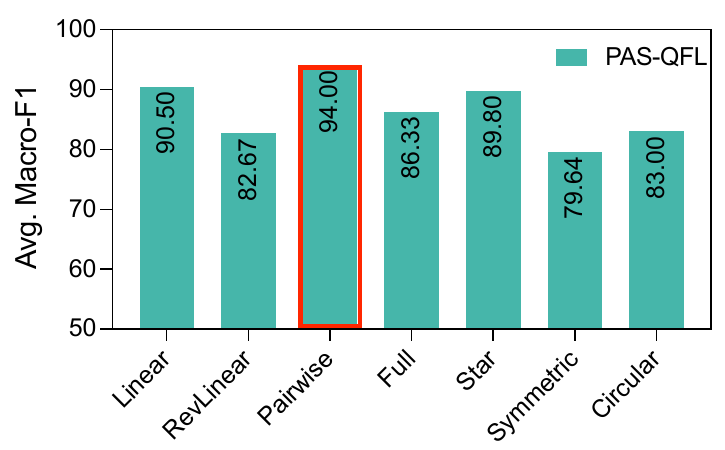}
\caption{Effect of shared ansatz selection on the FM-2 task with 5 clients. The personalized private ansatzes are fixed according to PAS-QFL, while the shared ansatz is varied across the candidate ansatz pool.}
\label{fig:9c_5q_share_selection}
\end{figure}

We next study the effect of shared ansatz selection. We use the same FM-2 setting with 5 clients and keep the personalized private ansatzes fixed according to PAS-QFL's local private selection. We then vary only the shared ansatz structure by replacing the selected shared ansatz with each candidate ansatz in the pool, while keeping the same federated training procedure. This setting isolates the impact of the shared ansatz from the effect of private ansatz personalization.

Fig.~\ref{fig:9c_5q_share_selection} shows that the choice of shared ansatz has a clear impact on the final federated performance. The shared ansatz selected by PAS-QFL, Pairwise, achieves the highest average Macro-F1 of 94.00. Other shared ansatzes lead to lower performance. This performance spread indicates that even when each client already uses a personalized private ansatz, the shared ansatz remains important because it defines the common component whose parameters are aggregated across clients.

These results support the role-specific design of PAS-QFL. The private ansatz improves client-specific local adaptation, but the shared ansatz must still provide a stable common structure for federated collaboration. Selecting the shared ansatz using the profiling-based stability-aware criterion helps avoid shared structures that are less compatible with cross-client aggregation, leading to stronger overall performance.

\section{Conclusion}
This paper presents PAS-QFL, a personalized ansatz selection framework for quantum federated learning under client data heterogeneity. 
By profiling candidate ansatzes across clients, PAS-QFL selects a stable shared ansatz for federated aggregation and a client-specific private ansatz that acts as a personalized quantum decision head, enabling ansatz-level personalization while keeping the aggregated parameters structurally consistent across clients. Experiments on Fashion-MNIST and MNIST, under both noiseless and noisy simulation, show that PAS-QFL improves average Macro-F1 over fixed-structure QFL baselines in most settings, with the clearest gains on the more challenging multi-class tasks. These results demonstrate that the structure of the private ansatz is an important personalization target in QFL, and that role-specific selection can improve local adaptation while preserving well-defined federated aggregation. Future work will extend PAS-QFL toward hardware-aware ansatz search and adaptive strategies that update the shared and private structures during training.

\bibliographystyle{IEEEtran}
\bibliography{ref}

\end{document}